\documentclass[aps,prl,amsmath,amssymb, twocolumn]{revtex4}
\usepackage[draftl]{changes}
\definechangesauthor[color=blue]{IB}
\definechangesauthor[color=orange]{GB}
\usepackage{units}

\usepackage[]{graphicx}
\usepackage{times}
\usepackage{bm}
\usepackage{color}
\renewcommand{\bm }{\mathbf}

\newcommand{\comment}[1]{}

\begin{document}

\title{Dark excitons in transition metal dichalcogenides}

\author{Ermin Malic$^{1}$, Malte Selig$^{2}$, Maja Feierabend$^{1}$, Samuel Brem$^{1}$, Dominik Christiansen$^{2}$, Florian Wendler$^{2}$, Andreas Knorr$^{2}$, Gunnar Bergh\"auser$^{1}$}
\email{gunbergh@chalmers.se}
\affiliation{$1$ Department of Physics, Chalmers University of Technology, Gothenburg, Sweden,
$2$ Institut f\"ur Theoretische Physik, Technische Universit\"at Berlin, Berlin, Germany
}

\begin{abstract}
Monolayer transition metal dichalcogenides (TMDs) exhibit a remarkably strong Coulomb interaction that manifests in tightly bound excitons. Due to the complex electronic band structure exhibiting several spin-split valleys in the conduction and valence band, dark excitonic states can be formed. They are inaccessibly by light due to the required spin-flip and/or momentum transfer. 
The relative position of these dark states with respect to the optically accessible bright excitons has a crucial impact on the emission efficiency of these materials and thus on their technological potential. Based on the solution of the Wannier equation, we present the excitonic landscape of the most studied TMD materials including the spectral position of momentum- and spin-forbidden excitonic states. We show that the knowledge of the electronic dispersion does not allow to conclude about the nature of the material's band gap, since excitonic effects can give rise to significant changes. Furthermore, we reveal that an exponentially reduced photoluminescence yield does not necessarily reflect a transition from a direct to a non-direct gap material, but can be ascribed in most cases to a change of the relative spectral distance between bright and dark excitonic states.

\end{abstract}

\maketitle

\section{Introduction}
\begin{figure}[t!]
\begin{center}
\includegraphics[width=1\linewidth]{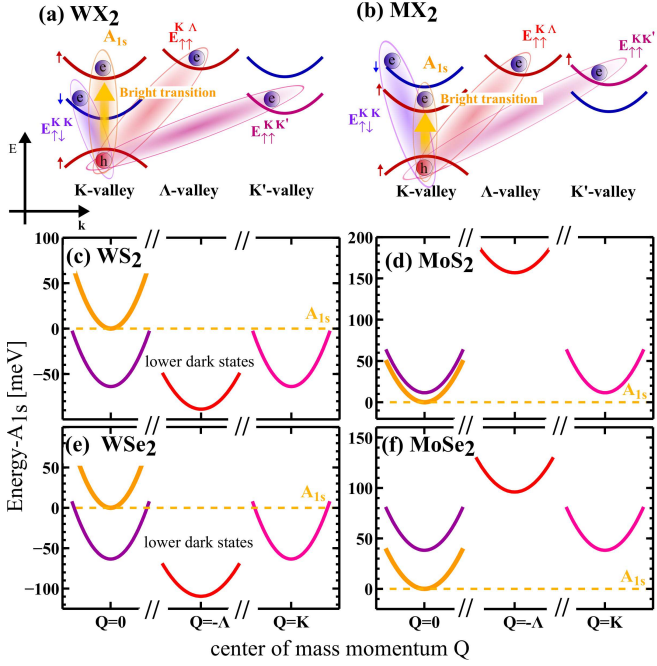}
\end{center}
\caption{\textbf{Dark and bright excitons with the hole located at the K valley.} Schematic electronic dispersions around the K and $\Lambda$ valley for (a) tungsten-based TMDs (WS$_2$ and WSe$_2$) and (b) molybdenum-based TMDs (MoS$_2$ and MoSe$_2$). In the first case, the lowest conduction and the highest valence band have the opposite spin. Spin-up and spin-down bands are denoted by red and blue lines, respectively. The spin-down valence band is not shown, since it is a few hundreds meV away contributing to B excitons that are not considered here. The yellow arrow describes the lowest optically induced transition between the bands of the same spin at the K point. The correlated electron-hole pairs are enclosed by a yellow (bright A$_{1s}$ exciton), red and orange (momentum-forbidden dark K-$\Lambda$ and K-K' exciton, respectively) and purple ovals (spin-forbidden dark K-K exciton). 
Exciton dispersion in (c) WS$_2$, (d) MoS$_2$, (e) WSe$_2$ and (f) MoSe$_2$ calculated by solving the Wannier equation. While in molybdenum-based TMDs, the bright exciton is the energetically lowest state (yellow line), tungsten-based TMDs exhibit lower lying dark excitonic states. 
}
 \label{fig1_Khole}
\end{figure}

\begin{figure}[t!]
\begin{center}
\includegraphics[width=1\linewidth]{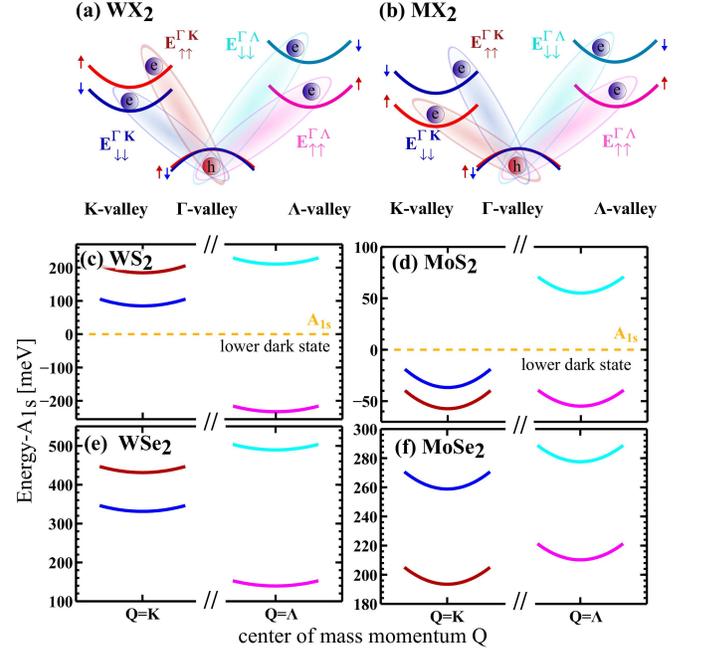}
\end{center}
\caption{\textbf{Dark and bright excitons with the hole located at the $\Gamma$ valley.} The same as in Fig. \ref{fig1_Khole}, however now considering $\Gamma-K$ and $\Gamma-\Lambda$ excitons, where the hole is located at the $\Gamma$ valley. Here, bright excitons do not appear, since there is no direct band gap at the $\Gamma$ point. Thus, we only find spin- and momentum-forbidden dark exciton states. 
Since there is no spin-orbit-induced splitting of the valence band at the $\Gamma$ point, for every spin-forbidden state there is a degenerate momentum-forbidden state.
We find that for TMDs including selen atoms ((e),(f)), $\Gamma$-hole excitons are located far above the lowest bright exciton transition (not shown), while TMDs including sulphur atoms ((c),(d)) exhibit excitons close to the bright state (dashed orange line). 
 }
 \label{fig2_Ghole}
\end{figure}

The complex electronic band structure combined with a strong spin-orbit coupling and the extremely efficient Coulomb interaction results in a remarkably versatile exciton landscape in monolayer transition metal dichalcogenides (TMDs) \cite{MacDonald2015PRB,Malte2016a,2017_Feierabend_NatComm,2017_Malte_Arxiv}. Besides the optically accessible Rydberg-like series of A$_{\rm{1s}}$, A$_{\rm{2s}}$, A$_{\rm{3s}}$, ... excitons, we also find a variety of dark states that cannot be excited by light \cite{Berkelbach2015PRB,MacDonald2015PRB,Dery2015PRB,Echeverry2016PRB,Potemski2017a,Danovich2017a,Steinhoff2014,2016_bergh_2Dmat,RudiandUs1}. The electronic band structure of TMDs exhibits four distinguished minima in the energetically lowest conduction band (K, K', $\Lambda$, and $\Lambda'$) and three maxima in the highest valence band (K, K', and $\Gamma$). Coulomb-bound electron-hole pairs can be formed within the K valley resulting in bright K-K excitons (yellow ovals in Figs. \ref{fig1_Khole}(a)-(b)). However, they can be also formed involving electrons and holes that are located in different valleys resulting in momentum-forbidden dark excitonic states (red and orange ovals in Figs. \ref{fig1_Khole}(a)-(b)). These states cannot be accessed by light, since photons cannot provide the required large center-of-mass momentum. Furthermore, we distinguish K-hole and $\Gamma$-hole states, where the hole is located either at the K valley (Fig. \ref{fig1_Khole}) or at the $\Gamma$ valley (Fig. \ref{fig2_Ghole}). The corresponding electron can then be either in the $\Lambda^{(\prime)}$ or the K$^{(\prime)}$ valley.

In addition to momentum-forbidden dark states, there is also a different class of dark excitons based on the spin. These so-called spin-forbidden (or spin-unlike) excitons consist of an electron and a hole with opposite spin and are optically inaccessible, since photons cannot induce a spin-flip process (purple oval in Figs. \ref{fig1_Khole}(a)-(b)). 
While the spin-orbit splitting in the valence band is as large as few hundreds of meV \cite{RamasubramaniamPRB2012,CheiwchanchamnangijPRB2012,2014_Zhang_NatNano}, the splitting of the conduction band is predicted to be much smaller in the range of few tens of meV \cite{Potemski2017a,Hongkun2017,Echeverry2016PRB,Barry_PRL_Louie2015}. 
Since the spin-orbit coupling can be positive or negative depending on the TMD material, two distinct orderings of spin states are possible: While in molybdenum-based TMDs (MoS$_2$ and MoSe$_2$), electrons in the lowest conduction band have the same spin as those in the highest valence band, an opposite spin ordering is found in tungsten-based TMDs (WS$_2$ and WSe$_2$) \cite{Kormanyos2014}, cf. Figs. \ref{fig1_Khole}(a) and (b).
While there have been many studies on electronic and excitonic properties in TMDs \cite{RamasubramaniamPRB2012,CheiwchanchamnangijPRB2012,RubioPRB2012,BerkelbachPRB2013,Chernikov2014,Bergh2014b}, the exciton landscape including bright as well as the variety of spin- and momentum-forbidden dark excitonic states has still not been entirely revealed 
\cite{Echeverry2016PRB,Dery2015PRB,Potemski2017a,MacDonald2015PRB,Barry_PRL_Louie2015}. In particular, the relative spectral position of bright and dark excitons is very important, since it determines whether the material is a direct- or an indirect-gap semiconductor, which has a crucial impact on the efficiency of light emission in TMDs \cite{HeinzPRL2010}. Thus, a profound knowledge about the exciton landscape in TMD materials is of high interest both for the fundamental science as well as technological application of these materials in future optoelectronic devices. 
In this work, we investigate the excitonic dispersion of different monolayer TMD materials including all different types of excitonic states. The approach is based on a numerical solution of the Wannier equation providing access to the full spectrum of exciton eigenenergies and eigenfunction \cite{Kochbuch,PQE,Bergh2014b, RudiandUs1,Malte2016a,2017_Feierabend_NatComm,2017_Malte_Arxiv}. The goal is to shed light on the relative spectral position of bright as well as momentum- and spin-forbidden dark excitonic states and investigate their impact on the photoluminescence (PL) quantum yield. Our main finding is that a drastic reduction of the yield (as observed in multilayer TMDs) does not automatically require a transition from a direct to a non-direct gap material. An increase of the relative spectral distance between bright and dark excitonic states of approximately \unit[100]{meV} already results in a decrease of the PL yield by two orders of magnitude. We show that the temperature behavior of the quantum yield is an excellent indicator for the position of dark excitonic states. 

\section{Theoretical approach}
We exploit the density matrix formalism \cite{Kochbuch,PQE}, which allows for a microscopic description of many-particle processes in TMDs on microscopic footing. The starting point is the solution of the Heisenberg equation of motion for single-particle quantities (singlets) 
$a^{\dagger}_i a_j$. Here, we have introduced the fermion operators $a_j$ and $a^{\dagger}_i$,
which annihilate and create a particle in the state $j$ and $i$, respectively. The applied many-particle Hamilton operator accounts for the free-particle contribution and the Coulomb interaction. 
The latter induces a well known many-particle hierarchy problem, which has been truncated on singlet level using the cluster expansion \cite{Kochbuch,PQE}. 
Since in this work, we are interested in linear optics, we can neglect the 
changes in carrier densities $f^{\lambda}_{i}=\langle a^{\dagger}_{i,\lambda} a_{j, \lambda} \rangle$ with the band index $\lambda=v,c$. The linear response of the material is determined by the microscopic polarization $p^{cv}_{ij}=\langle a^{\dagger}_{i, c} a_{j, v}\rangle$, which describes optically induced inter-band transitions. 
The later is dominated by bright excitons well below the quasi free particle band gap. For TMDs, it is of crucial importance to account for excitonic effects.

To calculate the exciton landscape in TMDs, we first use a separation ansatz allowing us to decouple the relative and the center-of-mass motion of Coulomb-bound electron-hole pairs. Similarly to the hydrogen problem, we introduce center-of-mass and relative momenta $\bf{Q}$ and $\bf{q}$, respectively. Here, $\bf{Q=k_{2}-k_{1}} $ and 
${\bf q}=\frac{m_{h_\mu}}{M^{\mu}}{\bf k_{1}}+\frac{m_{e_{\mu}}}{M^{\mu}} {\bf k_{2}}$ 
with the electron (hole) mass
$m_{e_\mu(h_\mu)}$ and the total mass $M^{\mu}=m_{e_{\mu}}+m_{h_{\mu}}$ of the carrier band index $\mu$. 
The later is a compound index including the electron ($\xi=K^{(\prime)}, \Lambda^{(\prime)}$) and 
hole ($\xi=K^{(\prime)}, \Gamma$) valleys and their spins ($s=\uparrow,\downarrow$).
 The center-of-mass momentum $\bm Q$ is determined by the difference of the momenta $\bm k_1, \bm k_2$ of the two bound particles. Here, we define $\bm Q$ with respect to the hole momentum, i.e. it gives the relative position of the electron in momentum space with respect to the hole.
The relative motion is determined by the total momentum of the bound electrons and holes. It can be described by the Wannier equation \cite{Kochbuch,PQE,kuhn04, Bergh2014b} reflecting the exciton eigenvalue problem and the homogeneous solution of the TMD Bloch equation for the microscopic polarization. The solution of the Wannier equation 
\begin{equation}\label{wannier}
 \varepsilon^\mu_{\bf q}
 \varphi_{\bf q}^{n_\mu} 
 -(1-f^{e_\mu}_{\bf{q}}-f^{h_\mu}_{\bf{q}}) \sum_{\bf k} 
 V_{c \mathbf k, v\mathbf q}^{c \mathbf q, v \bf k}
 \, \varphi_{\bf {q-k} }^{n_\mu}=E^{ n_\mu}\varphi_{\bf {q}}^{n_\mu}
\end{equation}
offers microscopic access to exciton eigenfunctions $\varphi_{\bf {q}}^{n_\mu}$ and eigenenergies $E^{n_\mu}$. The index $n_{\mu}$ describes different states (e.g. 1s, 2s, 3s .., 2p, 3p..) of the exciton $\mu$. Note that in the following we focus our study on the energetically lowest 1s states in different valleys. In the Wannier equation, we have introduced the attractive electron-hole contribution of the Coulomb interaction $V_{c \mathbf k, v\mathbf q}^{c \mathbf q, v \bf k}$ and the electron (hole) occupations $f^{e_\mu(h_\mu)}_{\bf{q}}$. The latter become important in doped TMD materials, where they also lead to a Coulomb-induced renormalization of the free-particle energy
 $ \varepsilon^{\mu}_{\bf q}=\frac{\hbar^2 q^{2}}{2m^{\mu}}$, where
 $m_{\mu}=(m_{h_{\mu}}+m_{e_{\mu}})/(m_{h_{\mu}} \, m_{e_{\mu}})$ is the reduced mass. 

The appearing Coulomb matrix elements are calculated within an effective Hamilton approach including the free-particle energy and the spin-orbit interaction \cite{Barry_PRL_Xiao2015}. 
The wave functions are expanded using plane waves and can be expressed as spinors equivalent to sublattices A and B
\begin{equation}
\label{effH}
\phi^{s\xi\lambda}(\mathbf k , \mathbf r)
=\dfrac{1}{\sqrt{A}}
\left( \begin{array}{c}C_{\text{A}}^{s\xi\lambda}(\mathbf{k})\\C_{\text{B}}^{s\xi\lambda}(\mathbf{k})\end{array} \right)
e^{i\mathbf k \cdot \mathbf r}
\end{equation}
with the wave function coefficients
$C_{l}^{s\xi\lambda}(\mathbf k)$. Here, we have introduced, the valley index $\xi=K, \Lambda, K'$, the spin index $s=\uparrow, \downarrow$, the band index $\lambda=c,v$ and the sample area $A$ that cancels out after performing the sum over the momentum.
The coefficients can be obtained by solving the Schr\"odinger equation for the free-particle Hamiltonian including the spin-orbit interaction. They read
for the valence band and the conduction bands 
$C_{A}^{s\xi v}(\mathbf k)=e^{-i\xi\phi}\cos(\gamma_{s\xi}(\mathbf k))$
, $C_{B}^{s \xi v}(\mathbf k)=-\sin(\gamma_{s\xi}(\mathbf k))$
and 
$C_{A}^{s\xi c}(\mathbf k)=e^{-i\xi}\sin(\gamma_{s\xi}(\mathbf k))$
, $C_{B}^{s\xi c}(\mathbf k)=\cos(\gamma_{s\xi}(\mathbf k))$
 with $\gamma_{s\xi}(k)=-\frac{\xi}{2}\arctan\big(2\hbar k [m_{s\xi}^{\lambda}\Delta_{cv}^{s\xi})]^{-\frac{1}{2}}\big)$.
These coefficients include the lattice-dependent symmetries and the resulting electronic band structure at the high symmetry points in the Brillouin zone of the investigated TMD material.
Thus, the effective Hamilton approach allows for a consistent description of all matrix elements, once the electronic band structure parameters, such as electronic band gaps, effective masses, and the spin-orbit coupling are known. The calculation of these parameters is beyond the scope of this work. 
Thus, for all our investigations we take a consistent and fixed set of input parameters from a DFT calculation by A. Korm\'anyos et al. \cite{Kormanyos2014}.

The general form of the Coulomb matrix elements reads
\begin{equation}
\label{coulomb}
 V^{ab}_{cd}=\sum_{q}V_{\mathbf q} \,\Gamma^{ac}_{\mathbf q}\Gamma^{bd}_{-\mathbf q}
\end{equation}
with the screened 2D Coulomb potential $V_{\mathbf q}$ and the factors 
$
\Gamma^{ij}_{\mathbf q}=\delta_{s_{i}s_{j}}\delta_{\bf q,\bf k_{i}-\bf k_{j}} \sum_l^{A,B}
C_{l}^{s_i\xi_i\lambda_i*}(\mathbf k_i)\,C_{l}^{s_j\xi_j\lambda_j}(\mathbf k_j)$
stemming from the lattice dependent overlap functions.
To account for the screening of the Coulomb interaction, we apply an effective Keldysh potential
$
 V_{\mathbf q}=\frac{1}{\varepsilon_0 \varepsilon_s \tilde{\varepsilon}_q
q},
$
which is known to describe well exciton properties in TMDs \cite{RubioPRB2012,BerkelbachPRB2013,Bergh2014b}. Here, $\varepsilon_0$ is the dielectric permittivity and $\varepsilon_s=(\varepsilon_1+\varepsilon_2)/2$ denotes the substrate-induced dielectric constant describing a TMD layer sandwiched between two media. Furthermore, we have adjusted the simple Keldysh screening $\varepsilon_q=1+r_0 q$ to advanced ab initio calculations \cite{Louie2016PRB}, where the screening is defined by the ratio of the full GW and the bare two-dimensional potential. We find that the DFT result can be fitted by the modified Keldysh potential with $\tilde{\varepsilon}_q=1+r_0q/\left({q}^{\frac{5}{3}}_{ } a_0^{\frac{5}{3}}+1\right)$. Here, the screening length $r_0=d\varepsilon_\bot/\varepsilon_s$ is determined by the thickness of the monolayer material $d \approx 7$ nm, $a_0\approx 0.3$ nm is the lattice constant and the TMD dependent dielectric tensor $\varepsilon_\bot=11.7$ for WS$_2$ and $\varepsilon_\bot=15.3$ for MoSe$_2$ \cite{Berkelbach2015PRB}. 

In this work, we have taken into account within the Wannier equation the attractive and repulsive Coulomb terms inducing formation of excitons and renormalization of electronic states, respectively. We have not considered the renormalization of the excitonic dispersion due to the Coulomb exchange interaction, which is known to lead to minor shifts of excitonic states in the range of a few meV \cite{Barry_PRL_Louie2015,MacDonald2015PRB,Echeverry2016PRB,HaiZhouPRLett2013,Ross2013NatCommun,
Hongyi_2014NCommValley2014,Steinhoff2017a}.
The actual strength of the Coulomb exchange coupling is still under debate in literature. While in \cite{Barry_PRL_Louie2015} for MoS$_2$ the renormalization induced by the exchange interaction is predicted to result in a strong blue-shift moving the spin-like K-K excitons energetically above the spin-unlike states, the calculations of \cite{Echeverry2016PRB} suggests that the renormalization via the exchange term is not sufficient to change the ordering of spin-like and spin-unlike states in TMDs.
Similar discrepancies in ab initio studies are also present for electronic band structure calculations.
 We would like to emphasize that the focus of our work does not lie on exact quantitative numbers for a specific TMD, but on revealing qualitative trends induced by the exciton landscape and their implications on experimentally accessible parameters, such as the photoluminescence quantum yield.

\section{Exciton landscape}

Evaluating the Wannier equation, we have full access to the eigenenergies of all exciton states. 
Figures \ref{fig1_Khole} (c)-(f) show the excitonic dispersion including the lowest A$_{\rm_{1s}}$ exciton states in the four most studied TMD materials (MoS$_2$, MoSe$_2$, WS$_2$, and WSe$_2$). We focus first on excitons, where the hole is located at the K point, while the electron can be either at the K, the $\Lambda$, or the K' point. We include all momentum- and spin-forbidden dark excitonic states. 
Interestingly, we find in tungsten-based TMDs dark excitons energetically below the bright K-K transition (orange dashed lines), i.e. these TMDs are indirect semiconductors, cf. Fig. \ref{fig1_Khole}(c),(e). 
The spin-forbidden K-K exciton, consisting of Coulomb-bound electrons and holes both located in the K valley but with the opposite spin, lie approximately \unit[50]{meV} below the bright state for WS$_2$ and WSe$_2$ on a SiO$_2$ substrate. 
Note that the momentum-forbidden spin-like K-K' excitons are energetically degenerate with the spin-forbidden K-K excitons, however they may become separated by a few meV through the Coulomb exchange coupling that only renormalizes spin-like states \cite{Hongyi_2014NCommValley2014,Barry_PRL_Louie2015,Echeverry2016PRB}. 

The lower energetic position of spin-forbidden states can be ascribed to the opposite ordering of spin states in tungsten-based TMDs (Fig. \ref{fig1_Khole}(a)). 
This is independent of the Coulomb interaction and can therefore not be tuned via screening of the Coulomb interaction. In contrast, the momentum-forbidden K-$\Lambda$ excitons consisting of electrons and holes in different valleys strongly depend on the strength of the Coulomb interaction and could principally be externally tuned via substrate-induced screening or doping. For the investigated case of undoped TMDs on the SiO$_2$ substrate at room temperature, we find the state 
\unit[35]{meV} (\unit[50]{meV}) below the bright state for WS$_2$ (WSe$_2$). The lower energetic position is due to the much larger effective mass of the $\Lambda$ valley (by a factor of 3) compared to the K valley. The effective mass directly enters the Wannier equation (Eq. (\ref{wannier})) giving rise to a larger exciton binding energy of K-$\Lambda$ excitons that compensates for the originally higher $\Lambda$ valley. In contrast, in molybdenum-based TMDs the energetic distance of the K and the $\Lambda$ valley is too high, so that these TMDs remain direct semiconductors also after considering excitonic effects (if only K-hole excitons are considered), cf. Figs. \ref{fig1_Khole}(d),(f).

Figure \ref{fig2_Ghole} shows the excitonic landscape of different TMDs now focusing on the excitons, where the hole is located at the $\Gamma$ point, while the electron is at the K, $\Lambda$, or K' valley. Note that the valence band at the $\Gamma$ point is not affected by the spin-orbit interaction resulting in double degeneracy at the $\Gamma$ point. Therefore, the spin-orbit coupling induced differences stem only from the conduction band.
Depending on how close the $\Gamma$ and the $\Lambda$ valley are to the conduction and the valence band at the K valley, we find dark states to be energetically higher (MoSe$_2$ and WSe$_2$) or lower (MoS$_2$ and WS$_2$) than the bright K-K exciton. The effective masses resulting from the lattice symmetry and spacial orbital overlaps are in all TMDs the smallest at the K point followed by the $\Lambda$ and the $\Gamma$ point. As a result, the exciton binding energies are the largest for electrons in the $\Lambda$ valley and holes occupying the $\Gamma$ point.
Furthermore, the excitonic dispersion of 
 $\Gamma$-hole excitons is much flatter compared to K-hole excitons, cf. Figs. \ref{fig1_Khole} and \ref{fig2_Ghole}.

\section{Quantum yield}

Having revealed the exciton landscape, we can now investigate the impact of dark states on the photoluminescence quantum yield in different TMD materials.
We estimate the yield $Y$ as the ratio of bright decay $\dot{N}_{\rm{bright}}|_{\rm{rad}}$ and total decay $\dot{N}_{\rm{tot}}$ this results in,
\begin{equation}
\label{yield}
Y=\frac{\gamma_{\text{rad}}N_{\rm{bright}}}{(\gamma_{\text{rad}}+\gamma_{\text{dark}})N_{\rm{bright}}+\gamma_{\text{dark}}N_{\rm{dark}}}
\end{equation} 
corresponding to the ratio of the thermalized population of bright excitons 
($N_{\rm{bright}}=\sum_{\mathbf Q}N^{A_{1s}}_{\bf Q}\delta_{\bf Q,\bf k_{pt}}$, where $\bf k_{pt}$ is the photon momentum, determined by the speed of light c and the photon frequency $k_{pt}\approx \frac{E_{A1s}}{\hbar c}$) weighted with the radiative decay $\gamma_{\text{rad}}\approx 1.6 $ meV 
 \cite{Moody2015} and the population of all excitonic states  including also dark excitons ($N_{\rm{dark}}=\sum_{\mathbf Q, \mu}N_{\bf Q}^\mu-N_{\rm{bright}}$). The population of the latter is weighted by the non-radiative decay
$\gamma_{\text{dark}}\approx 0.6$ meV 
 accounting for disorder-assisted relaxation channels 
 \cite{2015_Wang_Nletter}. Note that there are additional density-dependent non-radiative decay channels via Auger scattering \cite{2016_Danonvic_2DMat,2015_Wang_Nletter}, however, since we focus on the low excitation regime, these channels can be neglected in our work.
The above description of the quantum yield is a good estimate provided that the thermalization is much faster than the radiative decay \cite{2017_Malte_Arxiv,2000_Traenhard_PRB}.

\begin{figure}[t!]
\begin{center}
\includegraphics[width=1\linewidth]{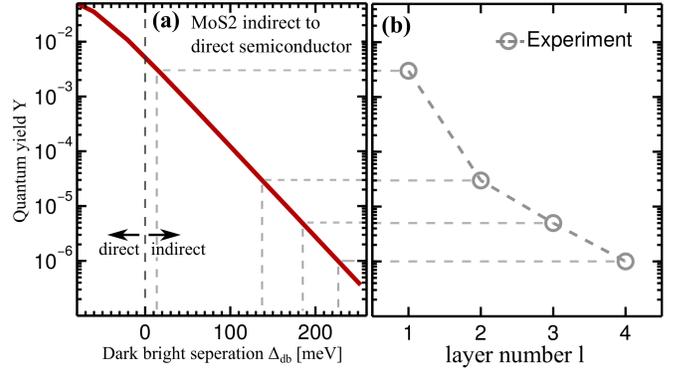}
\end{center}
\caption{\textbf{Impact of dark-bright separation and layer number on quantum yield.} 
 Quantum yield 
$Y$ as defined in Eq. (\ref{yield}) is shown in dependence of (a) the relative spectral position
of the bright and the lowest lying dark state  $\Delta_{db}=E_{\rm{bright}}-E_{\rm{dark}}$ and (b) the number of TMD layers $l$ in the case of MoS$_2$.
The quantum yield is extremely sensitive to $\Delta_{db}$ and shows an exponential dependence (except for the region around $\Delta_{db}\approx 0$) reflecting the Boltzmann distribution of excitonic states. Grey circles in (b) show experimental data on the layer dependence taken from Ref. \cite{HeinzPRL2010}. 
Using the experimental results we can estimate the dark-bright separation $\Delta^l_{db}$ for different layer numbers $l$. We find that monolayer MoS$_2$ is a slightly indirect-gap semiconductor with $\Delta^1_{db}\approx \unit[10]{meV}$. Note that the theoretical prediction based on the DFT input \cite{Kormanyos2014}for the electronic band structure gives a larger value of approximately $60$ meV.}
 \label{fig3}
 \end{figure}

\begin{figure}[t!]
\begin{center}
\includegraphics[width=1\linewidth]{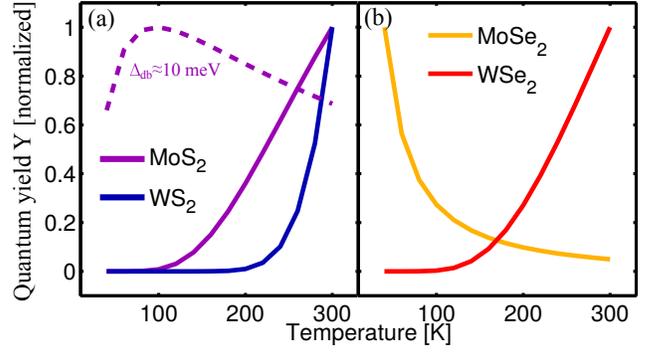}
\end{center}
\caption{\textbf{Temperature dependence of the quantum yield.} 
Quantum yield for monolayer (a) MoS$_2$, WS$_2$ and (b) MoSe$_2$, WSe$_2$ as a function of temperature. 
The relative position of dark and bright excitonic states in a TMD material can be directly read off by the temperature dependence of the quantum yield. 
An increasing (decreasing) quantum yield indicates a relatively higher (lower) occupation of the bright exciton state suggesting that the dark (bright) exciton is energetically lowest. 
While the solid lines are based on DFT input parameters on the electronic band structure \cite{Kormanyos2014}, the dashed line for MoS$_2$ shows the results obtained by assuming a dark-bright distance of $\Delta_{\text{db}}\approx \unit[10]{meV}$ according to the estimation from Fig. \ref{fig3}. In this case we find a non-monotonous temperature behavior, where after an initial increase the quantum yield starts to decrease again after a turning-point temperature of approximately \unit[100]{K}. This can be ascribed to the enhanced filling of dark states, which lie slightly above the bright A$_\text{1s}$ exciton (cf. Fig. \ref{fig1_Khole}(d)). 
}
 \label{fig4}
 \end{figure}

The time- and momentum-dependent exciton population $N^{n_\mu}_{\mathbf Q}(t)$ of the state $n_{\mu}$ can be obtained by taking into account the phonon-assisted formation of incoherent excitons as well as their thermalization towards an equilibrium distribution \cite{2017_Malte_Arxiv,2000_Traenhard_PRB}. Here, we assume a low-excitation limit, where the thermalized exciton populations follow the Boltzmann distribution. The exciton population within the light cone and with this the PL quantum yield are extremely sensitive to the relative spectral position of dark and bright states $\Delta_{db}^{\mu}=E_{\rm{bright}}-E_{\rm{dark}}^{\mu}$. In  Fig. \ref{fig3}(a) we have plotted the quantum yield in dependence on the lowest dark state $\mu_0$, i.e. $\Delta_{db}\equiv \Delta_{db}^{\mu_0}$. The smaller $\Delta_{db}$, the more excitons are located in the light cone and can decay radiatively enhancing the quantum yield by orders of magnitude. An important message of our work is that a transition from an indirect- to a direct-gap semiconductor is not necessary to explain a drastic increase in the quantum yield. Increasing $\Delta_{db}$ from 0 to \unit[100]{meV} leads to a change of more than two orders of magnitude in the quantum yield, cf. Fig. \ref{fig3}(a). The grey circles show the experimentally measured yield in dependence of the number of TMD layers, cf. Fig. \ref{fig3}(b). Our calculations show that the data can be explained by a change in the relative spectral position of dark and bright excitonic states, cf. the grey horizontal lines in Fig. \ref{fig3}. Here, the material does not necessarily need to become direct in the monolayer case. A clear experimental evidence for the direct or indirect nature of TMD materials can be obtained by measuring the temperature dependence of the photoluminescence quantum yield, as will be discussed in Fig. \ref{fig4}.

Now, we further evaluate the dependence of the quantum yield on $\Delta_{db}^{\mu}$.
Assuming a Boltzmann distribution for excitons, one finds an analytical expression for the quantum yield of TMD materials with a parabolic band structure
\begin{eqnarray}
\label{analytics}
Y&=&
\frac{\alpha\gamma_{\text{rad}} m_{\rm{A_{1s}}}
}{\alpha\gamma_{\text{rad}} m_{\rm{A_{1s}}}+
\gamma_{\text{dark}}\sum_{\mu}
m_{\mu}
e^{\beta\Delta_{db}^{\mu}}}
\\[7pt]\notag
&\approx&
\frac{\alpha\gamma_{\text{rad}}}{\gamma_{\text{dark}}}
\sum_{\mu}\frac{m_{\rm{A_{1s}}}}{m_{\mu}}
e^{-\beta\Delta_{db}^{\mu}}
 \end{eqnarray}
with $\beta=1/k_B T$ and $\alpha=\left(1-\exp{(-\frac{\beta \hbar^{2}k_{pt}^{2}}{2 m_{\rm{A_{1s}}}})}\right)
\approx \beta \dfrac{E_{A1s}^{2}}{2m_{A1s}c^2}$.  In the last step, we have assumed that the 
bright state lies well above the dark states, i.e. $\Delta^\mu_{db} \gg \beta^{-1}$. 
In this situation, we find an exponential decrease of the PL quantum yield with the relative dark-bright separation $\Delta^\mu_{db}$.
This explaines the linear dependence of the full solution of Eq.(\ref{yield}) for a growing dark-bright seperation $\Delta_{db}$ in the logarithmic plot of Fig. \ref{fig3} (a).
 We know from experiment that the quantum yield drastically decreases with the number of TMD layers \cite{HeinzPRL2010,2014_Zhang_NatNano,2013_Zeng_SciRep}. 
 We also know from ARPES and ab-initio studies that the relative position of K, $\Lambda$ and $\Gamma$ valleys shifts depending on the number of layers
 \cite{Jin2013,2014_Zhang_NatNano,2013_Zeng_SciRep}.
 
 With the obtained insights and by using the experimental results from Ref. \cite{HeinzPRL2010}, we can now estimate the relative shift of dark and bright excitons per additional layer in a TMD material, cf. the horizontal dashed grey lines in Fig.\ref{fig3}.
 Note that the dark-bright splitting $\Delta_{db}$ includes the energetic separation of the involved free-particle bands plus the changes in the excitonic binding energies.
 According to Fig. \ref{fig3}, we estimate the relative increase of dark-bright splitting from mono- to bilayer MoS$_2$ to be $\unit[140]{meV}$. This is in rather good agreement with first experimental data, where a relative shift of the involved electronic bands (K and $\Gamma$) is found to be in the range of \unit[400]{meV} 
 \cite{Jin2013}. 
 Taking into account the increased screening of the Coulomb interaction and the decreased effective masses of the involved electronic bands, we find that the binding energy for the most tigthly bound dark $\Gamma-\Lambda$ exciton reduces from approximately $800$ meV in monolayer MoS$_2$ to about $390$ meV in the bilayer case. At the same time, the bright A exciton reduces from roughly $460$ meV to $250$ meV. As a result, the measured relative shift of the involved electronic bands will be reduced by approximately 200 meV due to excitonic effects, which corresponds well to the predicted value for dark-bright splitting in the bilayer MoS$_2$.

Finally, we discuss the temperature dependence of the quantum yield. At \unit[0]{K}, all excitons occupy the lowest state. If this is a dark state, the quantum yield will be 0 and will then increase when the temperature rises reflecting the growing population of the spectrally higher bright states. If the lowest state is bright, then the quantum yield will be maximal at \unit[0]{K} and will then decrease at higher temperature. As a result, the temperature dependence is a clear indication for the nature of the band gap.
Figure \ref{fig4} shows the predicted quantum yield as a function of temperature for all investigated TMD materials. We find a clear increase of the yield for WS$_2$, WSe$_2$, and MoS$_2$, whereas the yield decreases for MoSe$_2$, i.e. only the latter is a direct semiconductor.

The increasing quantum yield of MoS$_2$ can be ascribed to the energetically lowest $\Gamma$-K exciton states, which are predominantly occupied at low temperatures. Note however that the used DFT input parameters \cite{Kormanyos2014} do not include substrate effects, which might modify orbital configurations. This could have a considerable impact in particular on the relative position of $\Gamma$ and $\Lambda$ valleys in the Brillouin zone \cite{Kormanyos2014}. Therefore, we also show
results for MoS$_2$, where we have used the dark-bright separation $\Delta_{\text{db}}\approx 10$ meV estimated from experimental data on the quantum yield (Fig.\ref{fig3}).
Here, the dark $\Gamma$-K excitons are expected to be only \unit[10]{meV} below the bright K-K exciton. As a result, at very low temperatures, we find the typical behavior for an indirect semiconductor, i.e. an increase in the quantum yield with temperature, cf. the dashed purple line in Fig. \ref{fig4} (a). However, above a certain temperature around \unit[100]{K}, we find a turning point and the quantum yield decreases again reflecting the behavior of a direct-gap semiconductor. 
This can be ascribed to the presence of dark states slightly above the bright exciton (Fig. \ref{fig1_Khole}(d)), which become filled resulting in a reduced occupation of the bright state and thus giving rise to a decrease in the quantum yield.
 Note that although MoS$_2$ is predicted to be an indirect-gap semiconductor, the temperature of the quantum yield shows the behavior that is expected from a direct-gap material - in agreement with experimental measurements from Ref. \cite{2015_Heinz_PRL}.

In summary, we provide a microscopic view on the exciton landscape in atomically thin transition metal dichalcogenides. Solving the Wannier equation we have full access to the spectral position of optically accessible bright excitons as well as momentum- and spin-forbidden dark excitonic states.
We show that neither a direct band gap in the electronic band structure of the material nor a drastic increase in the photoluminescence quantum yield are sufficient indicators for a direct gap semiconductor. We show that the experimentally observed increase in the quantum yield for monolayer TMDs does not necessarily reflect a transition from indirect to direct gap semiconductors, but can be explained by a change in the relative spectral position of bright and dark states. In conclusion, our work sheds light on the remarkably versatile exciton landscape of transition metal dichalcogenides and can guide future studies on these technologically promising materials.

We acknowledge financial support from the Swedish Research Council, the Stiftelsen Olle Engkvist, the Deutsche Forschungsgemeinschaft (DFG) through SFB 787, and the Chalmers Area of
Advance in Nanoscience and Nanotechnology. Furthermore, this project has also received funding from the European Unions Horizon 2020 research and innovation program 
under grant agreement No 696656 (Graphene Flagship) and from the project 734690 (SONAR).

\bibliographystyle{naturemag}

\begin{thebibliography}{10}
\expandafter\ifx\csname url\endcsname\relax
  \def\url#1{\texttt{#1}}\fi
\expandafter\ifx\csname urlprefix\endcsname\relax\def\urlprefix{URL }\fi
\providecommand{\bibinfo}[2]{#2}
\providecommand{\eprint}[2][]{\url{#2}}

\bibitem{MacDonald2015PRB}
\bibinfo{author}{Wu, F.}, \bibinfo{author}{Qu, F.} \&
  \bibinfo{author}{MacDonald, A.~H.}
\newblock \bibinfo{title}{{Exciton band structure of monolayer MoS$_{2}$}}.
\newblock \emph{\bibinfo{journal}{Phys. Rev. B}} \textbf{\bibinfo{volume}{91}},
  \bibinfo{pages}{075310} (\bibinfo{year}{2015}).

\bibitem{Malte2016a}
\bibinfo{author}{Selig, M.} \emph{et~al.}
\newblock \bibinfo{title}{{Excitonic linewidth and coherence lifetime in
  monolayer transition metal dichalcogenides}}.
\newblock \emph{\bibinfo{journal}{Nature Communications}}
  \textbf{\bibinfo{volume}{7}}, \bibinfo{pages}{13279} (\bibinfo{year}{2016}).

\bibitem{2017_Feierabend_NatComm}
\bibinfo{author}{Feierabend, M.}, \bibinfo{author}{Bergh{\"a}user, G.},
  \bibinfo{author}{Knorr, A.} \& \bibinfo{author}{Malic, E.}
\newblock \bibinfo{title}{Proposal for dark exciton based chemical sensors}.
\newblock \emph{\bibinfo{journal}{Nature Communications}}
  \bibinfo{pages}{14776} (\bibinfo{year}{2017}).

\bibitem{2017_Malte_Arxiv}
\bibinfo{author}{Selig, M.} \emph{et~al.}
\newblock \bibinfo{title}{Dark and bright exciton formation, thermalization,
  and photoluminescence in monolayer transition metal dichalcogenides}.
\newblock \emph{\bibinfo{journal}{arXiv:1703.03317}}  (\bibinfo{year}{2017}).

\bibitem{Berkelbach2015PRB}
\bibinfo{author}{Berkelbach, T.~C.}, \bibinfo{author}{Hybertsen, M.~S.} \&
  \bibinfo{author}{Reichman, D.~R.}
\newblock \bibinfo{title}{Bright and dark singlet excitons via linear and
  two-photon spectroscopy in monolayer transition-metal dichalcogenides}.
\newblock \emph{\bibinfo{journal}{Phys. Rev. B}} \textbf{\bibinfo{volume}{92}},
  \bibinfo{pages}{085413} (\bibinfo{year}{2015}).

\bibitem{Dery2015PRB}
\bibinfo{author}{Dery, H.} \& \bibinfo{author}{Song, Y.}
\newblock \bibinfo{title}{Polarization analysis of excitons in monolayer and
  bilayer transition-metal dichalcogenides}.
\newblock \emph{\bibinfo{journal}{Phys. Rev. B}} \textbf{\bibinfo{volume}{92}},
  \bibinfo{pages}{125431} (\bibinfo{year}{2015}).

\bibitem{Echeverry2016PRB}
\bibinfo{author}{Echeverry, J.~P.}, \bibinfo{author}{Urbaszek, B.},
  \bibinfo{author}{Amand, T.}, \bibinfo{author}{Marie, X.} \&
  \bibinfo{author}{Gerber, I.~C.}
\newblock \bibinfo{title}{Splitting between bright and dark excitons in
  transition metal dichalcogenide monolayers}.
\newblock \emph{\bibinfo{journal}{Phys. Rev. B}} \textbf{\bibinfo{volume}{93}},
  \bibinfo{pages}{121107} (\bibinfo{year}{2016}).

\bibitem{Potemski2017a}
\bibinfo{author}{Molas, M.~R.} \emph{et~al.}
\newblock \bibinfo{title}{Brightening of dark excitons in monolayers of
  semiconducting transition metal dichalcogenides}.
\newblock \emph{\bibinfo{journal}{2D Materials}} \textbf{\bibinfo{volume}{4}},
  \bibinfo{pages}{021003} (\bibinfo{year}{2017}).

\bibitem{Danovich2017a}
\bibinfo{author}{{Danovich Mark}}, \bibinfo{author}{{Z{\'o}lyomi Viktor}} \&
  \bibinfo{author}{{Fal{\rq}ko Vladimir I.}}
\newblock \bibinfo{title}{{Dark trions and biexcitons in WS$_2$ and WSe$_2$
  made bright by e-e scattering}}.
\newblock \emph{\bibinfo{journal}{Scientific Reports}}
  \textbf{\bibinfo{volume}{7}}, \bibinfo{pages}{45998} (\bibinfo{year}{2017}).

\bibitem{Steinhoff2014}
\bibinfo{author}{Steinhoff, A.}, \bibinfo{author}{Rosner, M.},
  \bibinfo{author}{Jahnke, F.}, \bibinfo{author}{Wehling, T.~O.} \&
  \bibinfo{author}{Gies, C.}
\newblock \bibinfo{title}{{Influence of Excited Carriers on the Optical and
  Electronic Properties of MoS$_2$}}.
\newblock \emph{\bibinfo{journal}{Nano Letters}} \textbf{\bibinfo{volume}{14}},
  \bibinfo{pages}{3743--3748} (\bibinfo{year}{2014}).

\bibitem{2016_bergh_2Dmat}
\bibinfo{author}{Bergh\"auser, G.}, \bibinfo{author}{Knorr, A.} \&
  \bibinfo{author}{Malic, E.}
\newblock \bibinfo{title}{Optical fingerprint of dark 2p-states in transition
  metal dichalcogenides}.
\newblock \emph{\bibinfo{journal}{2D Materials}} \textbf{\bibinfo{volume}{4}},
  \bibinfo{pages}{015029} (\bibinfo{year}{2017}).

\bibitem{RudiandUs1}
\bibinfo{author}{Schmidt, R.} \emph{et~al.}
\newblock \bibinfo{title}{Ultrafast coulomb-induced intervalley coupling in
  atomically thin ws2}.
\newblock \emph{\bibinfo{journal}{Nano Letters}} \textbf{\bibinfo{volume}{16}},
  \bibinfo{pages}{2945--2950} (\bibinfo{year}{2016}).

\bibitem{RamasubramaniamPRB2012}
\bibinfo{author}{Ramasubramaniam, A.}
\newblock \bibinfo{title}{Large excitonic effects in monolayers of molybdenum
  and tungsten dichalcogenides}.
\newblock \emph{\bibinfo{journal}{Phys. Rev. B}} \textbf{\bibinfo{volume}{86}},
  \bibinfo{pages}{115409} (\bibinfo{year}{2012}).

\bibitem{CheiwchanchamnangijPRB2012}
\bibinfo{author}{Cheiwchanchamnangij, T.} \& \bibinfo{author}{Lambrecht, W.
  R.~L.}
\newblock \bibinfo{title}{Quasiparticle band structure calculation of
  monolayer, bilayer, and bulk mos${}_{2}$}.
\newblock \emph{\bibinfo{journal}{Phys. Rev. B}} \textbf{\bibinfo{volume}{85}},
  \bibinfo{pages}{205302} (\bibinfo{year}{2012}).

\bibitem{2014_Zhang_NatNano}
\bibinfo{author}{{Zhang Yi}} \emph{et~al.}
\newblock \bibinfo{title}{{Direct observation of the transition from indirect
  to direct bandgap in atomically thin epitaxial MoSe2}}.
\newblock \emph{\bibinfo{journal}{Nat Nano}} \textbf{\bibinfo{volume}{9}},
  \bibinfo{pages}{111--115} (\bibinfo{year}{2014}).

\bibitem{Hongkun2017}
\bibinfo{author}{{Zhou You}} \emph{et~al.}
\newblock \bibinfo{title}{{Probing dark excitons in atomically thin
  semiconductors via near-field coupling to surface plasmon polaritons}}.
\newblock \emph{\bibinfo{journal}{Nat Nano}} \textbf{\bibinfo{volume}{advance
  online publication}} (\bibinfo{year}{2017}).

\bibitem{Barry_PRL_Louie2015}
\bibinfo{author}{Qiu, D.~Y.}, \bibinfo{author}{Cao, T.} \&
  \bibinfo{author}{Louie, S.~G.}
\newblock \bibinfo{title}{Nonanalyticity, valley quantum phases, and lightlike
  exciton dispersion in monolayer transition metal dichalcogenides: Theory and
  first-principles calculations}.
\newblock \emph{\bibinfo{journal}{Phys. Rev. Lett.}}
  \textbf{\bibinfo{volume}{115}}, \bibinfo{pages}{176801}
  (\bibinfo{year}{2015}).

\bibitem{Kormanyos2014}
\bibinfo{author}{Andor, K.} \emph{et~al.}
\newblock \bibinfo{title}{{k $\cdot$ p theory for two-dimensional transition
  metal dichalcogenide semiconductors}}.
\newblock \emph{\bibinfo{journal}{2D Materials}} \textbf{\bibinfo{volume}{2}},
  \bibinfo{pages}{022001} (\bibinfo{year}{2015}).

\bibitem{RubioPRB2012}
\bibinfo{author}{Cudazzo, P.}, \bibinfo{author}{Gatti, M.} \&
  \bibinfo{author}{Rubio, A.}
\newblock \bibinfo{title}{Plasmon dispersion in layered transition-metal
  dichalcogenides}.
\newblock \emph{\bibinfo{journal}{Phys. Rev. B}} \textbf{\bibinfo{volume}{86}},
  \bibinfo{pages}{075121} (\bibinfo{year}{2012}).

\bibitem{BerkelbachPRB2013}
\bibinfo{author}{Berkelbach, T.~C.}, \bibinfo{author}{Hybertsen, M.~S.} \&
  \bibinfo{author}{Reichman, D.~R.}
\newblock \bibinfo{title}{Theory of neutral and charged excitons in monolayer
  transition metal dichalcogenides}.
\newblock \emph{\bibinfo{journal}{Phys. Rev. B}} \textbf{\bibinfo{volume}{88}},
  \bibinfo{pages}{045318} (\bibinfo{year}{2013}).

\bibitem{Chernikov2014}
\bibinfo{author}{Chernikov, A.} \emph{et~al.}
\newblock \bibinfo{title}{{Exciton Binding Energy and Nonhydrogenic Rydberg
  Series in Monolayer WS$_2$}}.
\newblock \emph{\bibinfo{journal}{Physical Review Letters}}
  \textbf{\bibinfo{volume}{113}}, \bibinfo{pages}{76802}
  (\bibinfo{year}{2014}).

\bibitem{Bergh2014b}
\bibinfo{author}{Bergh{\"{a}}user, G.} \& \bibinfo{author}{Malic, E.}
\newblock \bibinfo{title}{{Analytical approach to excitonic properties of
  MoS$_{2}$}}.
\newblock \emph{\bibinfo{journal}{Phys. Rev. B}} \textbf{\bibinfo{volume}{89}},
  \bibinfo{pages}{125309} (\bibinfo{year}{2014}).

\bibitem{HeinzPRL2010}
\bibinfo{author}{Mak, K.~F.}, \bibinfo{author}{Lee, C.}, \bibinfo{author}{Hone,
  J.}, \bibinfo{author}{Shan, J.} \& \bibinfo{author}{Heinz, T.~F.}
\newblock \bibinfo{title}{{Atomically Thin MoS$_2$: A New Direct-Gap
  Semiconductor}}.
\newblock \emph{\bibinfo{journal}{Phys. Rev. Lett.}}
  \textbf{\bibinfo{volume}{105}}, \bibinfo{pages}{136805}
  (\bibinfo{year}{2010}).

\bibitem{Kochbuch}
\bibinfo{author}{Haug, H.} \& \bibinfo{author}{Koch, S.~W.}
\newblock \emph{\bibinfo{title}{Quantum Theory of the Optical and Electronic
  Properties of Semiconductors}} (\bibinfo{publisher}{World Scientific
  Publishing Co. Pre. Ltd.}, \bibinfo{year}{2004}), \bibinfo{edition}{fifth
  edit} edn.

\bibitem{PQE}
\bibinfo{author}{Kira, M.} \& \bibinfo{author}{Koch, S.~W.}
\newblock \bibinfo{title}{{Many-body correlations and excitonic effects in
  semiconductor spectroscopy}}.
\newblock \emph{\bibinfo{journal}{Progress in Quantum Electronics}}
  \textbf{\bibinfo{volume}{30}}, \bibinfo{pages}{155--296}
  (\bibinfo{year}{2006}).

\bibitem{kuhn04}
\bibinfo{author}{Axt, V.~M.} \& \bibinfo{author}{Kuhn, T.}
\newblock \bibinfo{title}{{Femtosecond spectroscopy in semiconductors: a key to
  coherences, correlations and quantum kinetics}}.
\newblock \emph{\bibinfo{journal}{Rep. Prog. Phys.}}
  \textbf{\bibinfo{volume}{67}}, \bibinfo{pages}{433} (\bibinfo{year}{2004}).

\bibitem{Barry_PRL_Xiao2015}
\bibinfo{author}{Zhou, J.}, \bibinfo{author}{Shan, W.-Y.},
  \bibinfo{author}{Yao, W.} \& \bibinfo{author}{Xiao, D.}
\newblock \bibinfo{title}{Berry phase modification to the energy spectrum of
  excitons}.
\newblock \emph{\bibinfo{journal}{Phys. Rev. Lett.}}
  \textbf{\bibinfo{volume}{115}}, \bibinfo{pages}{166803}
  (\bibinfo{year}{2015}).

\bibitem{Louie2016PRB}
\bibinfo{author}{Qiu, D.~Y.}, \bibinfo{author}{da~Jornada, F.~H.} \&
  \bibinfo{author}{Louie, S.~G.}
\newblock \bibinfo{title}{{Screening and many-body effects in two-dimensional
  crystals: Monolayer MoS$_{2}$}}.
\newblock \emph{\bibinfo{journal}{Phys. Rev. B}} \textbf{\bibinfo{volume}{93}},
  \bibinfo{pages}{235435} (\bibinfo{year}{2016}).

\bibitem{HaiZhouPRLett2013}
\bibinfo{author}{Lu, H.-Z.}, \bibinfo{author}{Yao, W.}, \bibinfo{author}{Xiao,
  D.} \& \bibinfo{author}{Shen, S.-Q.}
\newblock \bibinfo{title}{{Intervalley Scattering and Localization Behaviors of
  Spin-Valley Coupled Dirac Fermions}}.
\newblock \emph{\bibinfo{journal}{Phys. Rev. Lett.}}
  \textbf{\bibinfo{volume}{110}}, \bibinfo{pages}{16806}
  (\bibinfo{year}{2013}).

\bibitem{Ross2013NatCommun}
\bibinfo{author}{{Ross Jason S.}} \emph{et~al.}
\newblock \bibinfo{title}{{Electrical control of neutral and charged excitons
  in a monolayer semiconductor}}.
\newblock \emph{\bibinfo{journal}{Nat Commun}} \textbf{\bibinfo{volume}{4}},
  \bibinfo{pages}{1474} (\bibinfo{year}{2013}).

\bibitem{Hongyi_2014NCommValley2014}
\bibinfo{author}{Yu, H.}, \bibinfo{author}{Liu, G.-B.}, \bibinfo{author}{Gong,
  P.}, \bibinfo{author}{Xu, X.} \& \bibinfo{author}{Yao, W.}
\newblock \bibinfo{title}{{Dirac cones and Dirac saddle points of bright
  excitons in monolayer transition metal dichalcogenides}}.
\newblock \emph{\bibinfo{journal}{Nat Commun}} \textbf{\bibinfo{volume}{5}}
  (\bibinfo{year}{2014}).

\bibitem{Steinhoff2017a}
\bibinfo{author}{Steinhoff, A.} \emph{et~al.}
\newblock \bibinfo{title}{Excitons versus electron-hole plasma in monolayer
  transition metal dichalcogenide semiconductors}.
\newblock \emph{\bibinfo{journal}{arXiv:1705.05202}} .

\bibitem{Moody2015}
\bibinfo{author}{Moody, G.} \emph{et~al.}
\newblock \bibinfo{title}{{Intrinsic Exciton Linewidth in Monolayer Transition
  Metal Dichalcogenides}}.
\newblock \emph{\bibinfo{journal}{Nature Commun.}}
  \textbf{\bibinfo{volume}{6}}, \bibinfo{pages}{8315} (\bibinfo{year}{2015}).

\bibitem{2015_Wang_Nletter}
\bibinfo{author}{Wang, H.}, \bibinfo{author}{Zhang, C.} \&
  \bibinfo{author}{Rana, F.}
\newblock \bibinfo{title}{Ultrafast dynamics of defect-assisted electron–hole
  recombination in monolayer mos2}.
\newblock \emph{\bibinfo{journal}{Nano Letters}} \textbf{\bibinfo{volume}{15}},
  \bibinfo{pages}{339--345} (\bibinfo{year}{2015}).
\newblock \bibinfo{note}{PMID: 25546602}.

\bibitem{2016_Danonvic_2DMat}
\bibinfo{author}{Danovich, M.}, \bibinfo{author}{Zólyomi, V.},
  \bibinfo{author}{Fal’ko, V.~I.} \& \bibinfo{author}{Aleiner, I.~L.}
\newblock \bibinfo{title}{Auger recombination of dark excitons in ws 2 and wse
  2 monolayers}.
\newblock \emph{\bibinfo{journal}{2D Materials}} \textbf{\bibinfo{volume}{3}},
  \bibinfo{pages}{035011} (\bibinfo{year}{2016}).

\bibitem{2000_Traenhard_PRB}
\bibinfo{author}{Thr\"anhardt, A.}, \bibinfo{author}{Kuckenburg, S.},
  \bibinfo{author}{Knorr, A.}, \bibinfo{author}{Meier, T.} \&
  \bibinfo{author}{Koch, S.~W.}
\newblock \bibinfo{title}{Quantum theory of phonon-assisted exciton formation
  and luminescence in semiconductor quantum wells}.
\newblock \emph{\bibinfo{journal}{Phys. Rev. B}} \textbf{\bibinfo{volume}{62}},
  \bibinfo{pages}{2706--2720} (\bibinfo{year}{2000}).

\bibitem{2013_Zeng_SciRep}
\bibinfo{author}{{Zeng Hualing}} \emph{et~al.}
\newblock \bibinfo{title}{{Optical signature of symmetry variations and
  spin-valley coupling in atomically thin tungsten dichalcogenides}}
  \textbf{\bibinfo{volume}{3}}, \bibinfo{pages}{1608} (\bibinfo{year}{2013}).

\bibitem{Jin2013}
\bibinfo{author}{Jin, W.~C.} \emph{et~al.}
\newblock \bibinfo{title}{{Direct Measurement of the Thickness-Dependent
  Electronic Band Structure of MoS$_2$ Using Angle-Resolved Photoemission
  Spectroscopy}}.
\newblock \emph{\bibinfo{journal}{Physical Review Letters}}
  \textbf{\bibinfo{volume}{111}} (\bibinfo{year}{2013}).

\bibitem{2015_Heinz_PRL}
\bibinfo{author}{Zhang, X.-X.}, \bibinfo{author}{You, Y.},
  \bibinfo{author}{Zhao, S. Y.~F.} \& \bibinfo{author}{Heinz, T.~F.}
\newblock \bibinfo{title}{{Experimental Evidence for Dark Excitons in Monolayer
  WSe$_{2}$}}.
\newblock \emph{\bibinfo{journal}{Phys. Rev. Lett.}}
  \textbf{\bibinfo{volume}{115}}, \bibinfo{pages}{257403}
  (\bibinfo{year}{2015}).

\end{thebibliography}

\end{document}